\documentclass{aa}
\usepackage[varg]{txfonts} 
\usepackage{courier} 
\usepackage{bm} 
\usepackage{upgreek} 
\usepackage[table,  dvipsnames]{xcolor} 
\usepackage{amsmath,amsfonts,amssymb} 
\usepackage{mathtools} 
\usepackage{hyperref} 
\usepackage{subfig} 
\usepackage{float} 
\usepackage{nicefrac} 

\hypersetup{colorlinks=true,citecolor=blue}

\usepackage{graphicx}  
\graphicspath{{../python/pics/}}

    \usepackage{etoolbox}
    \makeatletter
    
    \patchcmd{\NAT@citex}
      {\@citea\NAT@hyper@{%
         \NAT@nmfmt{\NAT@nm}%
         \hyper@natlinkbreak{\NAT@aysep\NAT@spacechar}{\@citeb\@extra@b@citeb}%
         \NAT@date}}
      {\@citea\NAT@nmfmt{\NAT@nm}%
       \NAT@aysep\NAT@spacechar\NAT@hyper@{\NAT@date}}{}{}
    
    \patchcmd{\NAT@citex}
      {\@citea\NAT@hyper@{%
         \NAT@nmfmt{\NAT@nm}%
         \hyper@natlinkbreak{\NAT@spacechar\NAT@@open\if*#1*\else#1\NAT@spacechar\fi}%
           {\@citeb\@extra@b@citeb}%
         \NAT@date}}
      {\@citea\NAT@nmfmt{\NAT@nm}%
       \NAT@spacechar\NAT@@open\if*#1*\else#1\NAT@spacechar\fi\NAT@hyper@{\NAT@date}}
      {}{}
    \makeatother

\newcommand{\D}[2]{\frac{\mathrm{d}{#1}}{\mathrm{d}{#2}}}

\newcommand{\filler}{--}
\begin{document}

\title{Importance of radiative effects in gap opening\\by planets in protoplanetary disks}

\author{
        Alexandros Ziampras\inst{\ref{inst1}}
        \and Wilhelm Kley\inst{\ref{inst1}}
        \and Cornelis P. Dullemond\inst{\ref{inst2}}
}

\institute{
        Institut f{\"u}r Astronomie und Astrophysik, Universit{\"a}t T{\"u}bingen, Auf der Morgenstelle 10, 72076 T{\"u}bingen, Germany\label{inst1}
        \and Institute for Theoretical Astrophysics, Heidelberg University, Albert-Ueberle-Str. 2, 69120 Heidelberg, Germany\label{inst2}
}

\date{\today}

\abstract
        {
                Recent ALMA observations revealed concentric annular structures in several young class-II objects. In an attempt to produce the rings and gaps in some of these systems, they have been modeled numerically with a single embedded planet assuming a locally isothermal equation of state. This is often justified by observations targeting the irradiation-dominated outer regions of disks (approximately~100\,au). We test this assumption by conducting hydrodynamics simulations of embedded planets in thin locally isothermal and radiative disks that mimic the systems HD~163296 and AS~209 in order to examine the effect of including the energy equation in a seemingly locally isothermal environment as far as planet--disk interaction is concerned.\\We find that modeling such disks with an ideal equation of state makes a difference in terms of the number of produced rings and the spiral arm contrast in the disk. Locally isothermal disks produce sharper annular or azimuthal features and overestimate a single planet's gap-opening capabilities by producing multiple gaps. In contrast, planets in radiative disks carve a single gap for typical disk parameters. Consequently, for accurate modeling of planets with semimajor axes up to about 100\,au, radiative effects should be taken into account even in seemingly locally isothermal disks. In addition, for the case of AS~209, we find that the primary gap is significantly different between locally isothermal and radiative models. Our results suggest that multiple planets are required to explain the ring-rich structures in such systems.
}


\bibpunct{(}{)}{;}{a}{}{,}

        \maketitle
        \section{Introduction}
        \label{section:intro}
        
        Planets are born in protoplanetary disks. While they do not always make their presence clear like in the case of PDS~70b and c \citep{keppler-etal-2018, haffert-etal-2019}, the ALMA observations of the DSHARP survey \citep{andrews-etal-2018} have provided theorists with high-fidelity datasets for testing the planet--disk interaction theory. It is fascinating how well a single planet can reproduce the annular substructures in some of the observed systems, for instance,~\citet{zhang-etal-2018}.
        
        Most of the aforementioned systems show resolved features at the 100au scale, suggesting that the disk temperature profile is set by a balance between heating by stellar irradiation and thermal cooling. This makes modeling these systems with a locally isothermal equation of state quite attractive because the cooling timescale in the optically thin irradiation-dominated outer disk is commonly sufficiently short (shorter than a hundredth of an orbit) to render radiative effects negligible. At the same time, the very low level of effective viscosity that is inferred for these disks \citep{zhang-etal-2018} indicates minuscule contributions by viscous heating to the thermal budget of the disk, such that the locally isothermal assumption seems further justified.
        
        However, an embedded planet also interacts with the disk gravitationally, forming spiral arms \citep{ogilvie-lubow-2002,rafikov-2002} that permeate the disk and can steepen into shocks \citep{zhu-etal-2015}. Several studies have shown that heating by these spiral shocks is another significant heat source at distances of several au for solar-type stars \citep{richert-etal-2015,lyra-etal-2016,rafikov-2016,ziampras-etal-2020}. While the planet's contribution to the energy budget decreases with increasing distances from the star, there is a gray area at a few tens of au where shock heating could operate to some degree, even when the cooling timescale is a small fraction of the local orbital period.

        In addition, \citet{miranda-rafikov-2019a} showed that even in seemingly locally isothermal scenarios, assuming an adiabatic equation of state results in fundamentally different physics concerning the angular momentum flux in protoplanetary disks, and it therefore leads to noticeable changes in dust continuum profiles. They quoted distances of about 80\,au as the lower limit beyond which a locally isothermal equation of state can indeed be justified, and showed that deviations of about $10^{-3}$ from $\gamma=1$ can lead to noticeable differences (of about 10\%) in simulated outcomes and therefore observables such as continuum emission intensity.
        
        This subject is quickly gaining clarity as more effort is made to understand planet-induced gap opening in order to interpret observations and constrain the properties of protoplanetary disks. During the reviewing process for this paper, two publications by \citet{miranda-rafikov-2019c} and \citet{zhang-zhu-2019} were made available, which discuss the effect of a finite cooling timescale on the location and number of gaps that can be produced by a single planet. Our study does not reiterate these new results, but rather enriches them by handling radiative effects differently. It serves as an additional point of view in arriving at a similar conclusion: radiative cooling is of key importance in determining the radial structure of a disk with an embedded planet.
        
        We use numerical simulations to address the importance of proper treatment of radiative effects for two systems that mimic the properties of HD~163296 and AS~209 \citep{andrews-etal-2018,zhang-etal-2018} and show that an adiabatic equation of state can produce different results with respect to a locally isothermal model when the gap-opening capabilities of a planet are modeled.
        \begin{figure*}[h]
                \centering
                \begin{minipage}{0.48\textwidth}
                        \includegraphics[width=\textwidth]{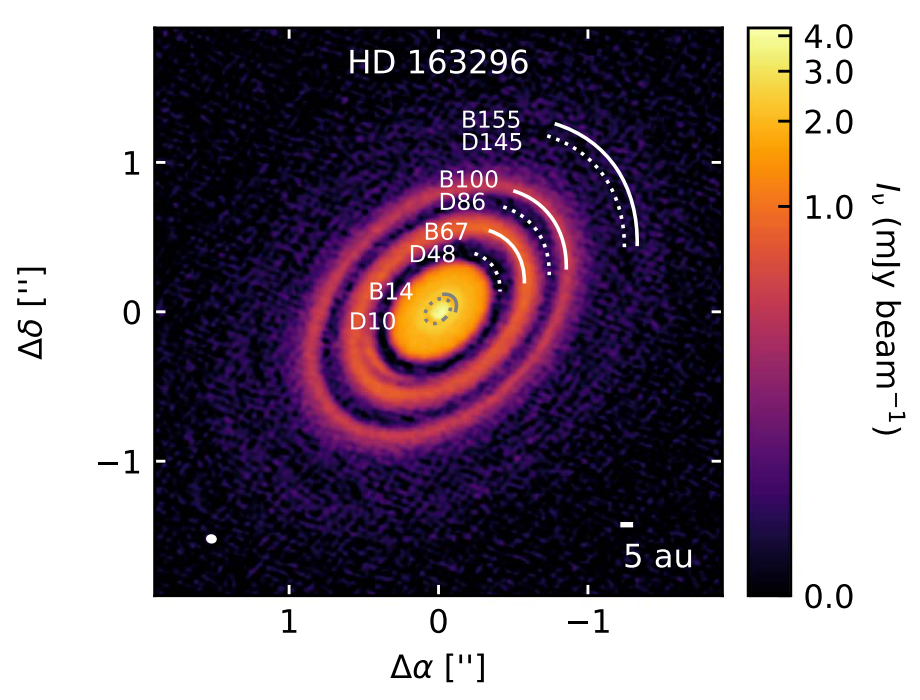}
                \end{minipage}
                \begin{minipage}{0.48\textwidth}
                        \includegraphics[width=\textwidth]{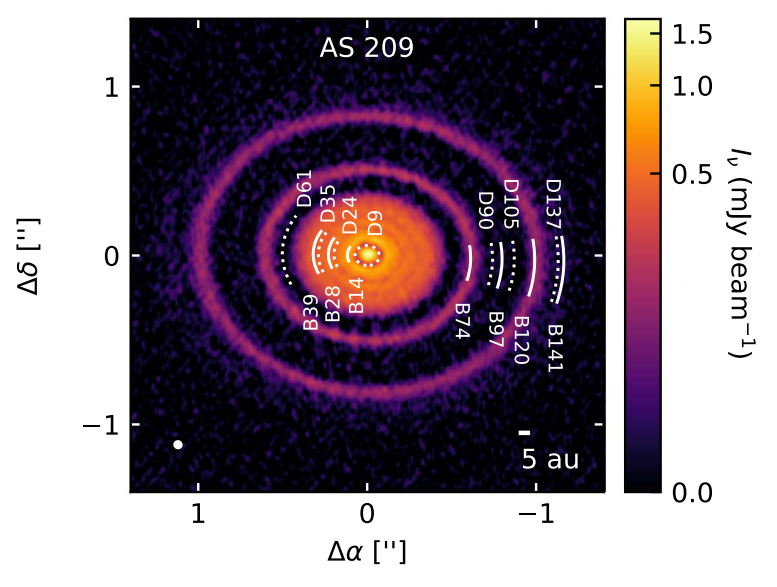}
                \end{minipage}
                \caption{ALMA continuum emission observations of the two systems we used as testbeds for our numerical models, as shown in \citet{huang-etal-2018}. Solid and dotted white arcs mark the location of bright rings and dark gaps, respectively, along with their distance from the host star in au. It is likely that a planet is responsible for one or more of the gaps observed in either system. Left: HD~163296, a 12.6 Myr old likely isolated system that features at least two clearly visible gaps at 48 and 86\,au. Our models focus on the 20--60\,au range, with a planet carving a gap at 48\,au. Right: AS~209, a 1 Myr old system in the Ophiuchus region that is rich in annular structures, boasting five rings between 20--130\,au. This source has been modeled by \citet{zhang-etal-2018}, who found that a fit with a single planet at 99\,au reconstructed all five of these rings.}
                \label{fig:dsharp}
        \end{figure*}

        In Section~\ref{section:physics-numerics} we present our physical and numerical setup in modeling these two systems. We present our results in terms of disk structures in Section~\ref{section:results}, and discuss their implications in Section~\ref{section:discussion}. Finally, we summarize our results and conclude this study in Section~\ref{section:conclusions}.
        
        \section{Sources, physics, and numerics}
        \label{section:physics-numerics}
        
        In this section we describe the physical and numerical modeling of HD~163296 and AS~209. We first present the two systems that we use as our reference. We then describe the source terms in the vertically integrated hydrodynamics equations, our physical assumptions in terms of star, planet, and disk properties, and the initial and boundary conditions for our simulations.

        \subsection{Sources: HD~163296 and AS~209}
        Rather than carrying out a study on an arbitrary toy model of a protoplanetary disk, we decided to model two of the sources targeted by the DSHARP survey \citep{andrews-etal-2018}, namely HD~163296 and AS~209 (see Fig.~\ref{fig:dsharp}). Both of these sources feature annular structures in the form of bright rings and dark gaps in dust continuum emission images, with the former also showing a crescent at roughly 50\,au and the latter standing out with its ring-rich emission profile. These systems have been modeled in the past and their various features have been studied \citep{huang-etal-2018, dullemond-etal-2018, zhang-etal-2018, zhang-zhu-2019}, and using them as testbeds for our study can provide a deeper insight on any planets they might harbor.
        
        \subsection{Hydrodynamics, heating, and cooling}
        \label{section:physics}
        
        The two-dimensional (2D) vertically integrated Navier--Stokes equations in cylindrical $\{r,\phi,z\}$ coordinates for a perfect gas with surface density $\Sigma$, velocity $\bm{v}$ and specific midplane internal energy $e$ read
        \begin{subequations}
                \label{eq:navier-stokes}
                \begin{align}
                        \label{eq:navier-stokes-1}
                        \D{\Sigma}{t}&=-\Sigma\nabla\cdot\bm{v},
                \end{align}
                \begin{align}
                \label{eq:navier-stokes-2}
                \Sigma\D{\bm{v}}{t}&=-\nabla p+\Sigma \bm{g}+\nabla\cdot\bm{\upsigma},
                \end{align}
                \begin{align}
                \label{eq:navier-stokes-3}
                \D{\Sigma e}{t}&=-\gamma\Sigma e\nabla\cdot\bm{v}+Q_\mathrm{visc}+Q_\mathrm{irr}-Q_\mathrm{cool},
                \end{align}
        \end{subequations}
        where $\gamma$ is the adiabatic index and $p=(\gamma-1)\Sigma e$ is the vertically integrated pressure. External source terms (in our case, gravitational forces) are contained in $\bm{g}$, and $\bm{\upsigma}$ denotes the viscous stress tensor \citep{tassoul-1978}.
        
        For a thin disk of gas rotating on a Keplerian orbit with orbital frequency $\Omega_\mathrm{K} = \sqrt{\textrm{G}M/r^3}$, at distance $r$ around a star of mass $M$, the pressure scale height of the gas is $H={c_\mathrm{s}}_\mathrm{iso}/\Omega_\mathrm{K}$. Here, 
        ${c_\mathrm{s}}_\mathrm{iso}=\sqrt{p/\Sigma}$ is the isothermal sound speed and relates to the adiabatic sound speed $c_\mathrm{s}$ as $c_\mathrm{s}=\sqrt{\gamma}{c_\mathrm{s}}_\mathrm{iso}$. The aspect ratio of the disk is then $h=H/r$. The gravitational constant and the mean molecular weight of the gas are denoted by $\mathrm{G}$ and $\mu$, respectively. We adopted a typical solar-composition disk of molecular H-He gas, therefore $\gamma=\nicefrac{7}{5}$ and $\mu=2.35$.
        
        In a locally isothermal framework, the energy equation in Eq.~\eqref{eq:navier-stokes} is not solved. Instead, a fixed sound speed profile is adopted. This in turn defines a (fixed) radial temperature profile $T$, as ${c_\mathrm{s}}_\mathrm{iso}=\sqrt{\mathrm{R_{gas}}T/\mu}$, with $\mathrm{R_{gas}}$ being the gas constant. In the locally isothermal scenario this temperature profile is constant on cylinders and only dependent on radius, but we focus on the midplane temperature in this 2D approximation.
                
        More generally, however, the full system of equations is solved. In this case, we include viscous heating, stellar irradiation, and thermal cooling in the energy equation as follows:
        \begin{subequations}
        \label{eq:source-terms}
        \begin{align}
        \label{eq:source-terms-1}
        Q_\mathrm{visc}&
        =\frac{1}{2\nu\Sigma}\mathrm{Tr}(\bm{\upsigma}^2)=\frac{1}{2\nu\Sigma}\left(\sigma_{rr}^2+2\sigma_{r\phi}^2+\sigma_{\phi\phi}^2+\sigma_{zz}^2\right)
        \end{align}
        \begin{align}
        \label{eq:source-terms-2}
        Q_\mathrm{irr}&=2\frac{L_\ast}{4\mathrm{\pi} r^2}(1-\epsilon)\left(\D{\log H}{\log r}-1\right)h\frac{1}{\tau_\mathrm{eff}},
        \end{align}
        \begin{align}
        \label{eq:source-terms-3}
        Q_\mathrm{cool}&=2\sigma_\mathrm{SB}\frac{T^4}{\tau_\mathrm{eff}},
        \end{align}
        \end{subequations}
        where $\nu$ is the kinematic viscosity. For this we used the $\alpha$-ansatz \citep{shakura-sunyaev-1973}, which for thin disks reads $\nu=\alpha c_\mathrm{s} H$. The Stefan-Boltzmann constant is denoted by $\sigma_\mathrm{SB}$, and $\tau_\mathrm{eff}$ is an effective optical depth following \citet{hubeny-1990}
        \begin{equation}
        \tau_\mathrm{eff}=\frac{3\tau}{8}+\frac{\sqrt{3}}{4}+\frac{1}{4\tau},
        \end{equation}

        where $\tau$ is the vertical optical depth measured from the midplane to the disk surface,
        \begin{equation}
                \tau=\int_{0}^{\infty}\kappa\rho dz\approx c_1\kappa_\text{mid}\rho_\mathrm{mid}H.
        \end{equation}

        Here, $\rho$ is the volume density and $\kappa_\text{mid}(\rho,T)$ is the Rosseland mean opacity at the disk midplane, which we refer to as simply $\kappa$ and evaluate in Sect.~\ref{section:numerics} below, either using a constant value or adopting the temperature-dependent opacity law.
        To match the opacity drop with height, we included a correction factor~$c_1=\nicefrac{1}{2}$ \citep{mueller-kley-2011}. The midplane gas density $\rho_\mathrm{mid}$ is related to the surface density such that $\Sigma=\sqrt{2\mathrm{\pi}}\rho_\mathrm{mid}H$ (i.e., corresponding to a Gaussian vertical stratification at hydrostatic equilibrium).
        
        Our cooling and irradiation prescription adapts the model by \citet{menou-goodman-2004} for a disk with an albedo of $\epsilon=\nicefrac{1}{2}$ around a star with luminosity $L_\ast$. The factor $\D{\log H}{\log r}$ was assumed to be constant and equal to $\nicefrac{9}{7}$ (i.e., disk self-shadowing was not considered).
        
		For more details, see the physical setup by \citet{ziampras-etal-2020}.

        \subsection{Numerics}
        \label{section:numerics}
        
        We used the \texttt{PLUTO} code \citep{mignone-etal-2007} along with the FARGO method \citep{masset-2000,mignone-etal-2012} for our simulations. Models with an embedded planet were run on a polar $\{r,\phi\}$ grid, logarithmically spaced in the radial direction.
        
        In order to set our initial conditions, we constructed two different disks that mimic the structure of HD~163296 and AS~209. We therefore set $M_\ast$ and $L_\ast$ according to \citet{andrews-etal-2018}, and decided to embed a single planet with mass $M_\mathrm{p}=0.5$\,$\mathrm{M_J}$ at $r_\mathrm{p} = 48\,$au for HD~163296 and  $M_\mathrm{p}=0.083$\,$\mathrm{M_J}$ at $r_\mathrm{p} = 99\,$au for AS~209. We adopted a viscous $\alpha$ following \citet{zhang-etal-2018} (either $10^{-4}$ or $10^{-5}$, constant throughout the disk). For the Rosseland mean opacity we chose either a constant value $\kappa = 0.45\,\mathrm{cm}^2/\mathrm{g}$ (adapted from \citet{birnstiel-etal-2018}) or a temperature-dependent opacity law by \citet{lin-papaloizou-1985}. The latter dictates that $\kappa=5\times 10^{-4} T^2$ for temperatures under 170\,K, which is true for the full extent of the simulated disk. A list of the physical and numerical parameters used in our models is given in Table~\ref{table:HD-VS-AS}.
        
        The initial surface density needs to be prescribed, and the temperature is simply given by $Q_\mathrm{visc}+Q_\mathrm{irr} = Q_\mathrm{cool}$ (see Eq.~\eqref{eq:navier-stokes-3}) in the absence of a planet (we can ignore compression heating in this scenario because its contribution is negligible). We adopted values inferred in Table 3 by \citet{zhang-etal-2018} for HD~163296 such that the surface density at $r_\mathrm{gap}=\{10,48,86\}\,$au is $\Sigma(r_\mathrm{gap})=\{100,10,3\}\,\mathrm{g/ cm}^{2}$, respectively. We then fit a power law to these three points and find that a fit of
        \begin{equation}
                \label{eq:initial-surface-density-HD}
                \Sigma(r)^\mathrm{HD} \approx 10\,\left(\frac{r}{r_\mathrm{p}}\right)^{-3/2} \frac{\text{g}}{\text{cm}^2} \approx    3200\,\left(\frac{r}{\mathrm{au}}\right)^{-3/2} \frac{\text{g}}{\text{cm}^2}
        \end{equation}
        matches very well. For AS~209, we simply adopt a setup similar to that by \citet{zhang-etal-2018}, such that
        \begin{equation}
        \label{eq:initial-surface-density-AS}
        \Sigma(r)^\mathrm{AS} \approx 10\,\left(\frac{r}{r_\mathrm{p}}\right)^{-1} \frac{\text{g}}{\text{cm}^2}
        .\end{equation}

        As far as temperature is concerned, the effect of viscous heating $Q_\mathrm{visc}$ is functionally negligible within our simulation range for our $\alpha$ value. In the absence of a planet, the temperature profile is therefore set by a balance between irradiation heating and thermal cooling. In Eqs.~\eqref{eq:source-terms-2}~and~\eqref{eq:source-terms-3}, the optical depths for absorption and emission in $Q_\mathrm{irr}$ and $Q_\mathrm{cool}$ , respectively, can in principle be different, but as a first approximation we assumed that they are identical and can be factored out when we set $Q_\mathrm{irr} = Q_\mathrm{cool}$. The temperature profile then only depends on stellar properties so that

        \begin{equation}
                \label{eq:initial-temperature}
                T(r) = T_\mathrm{p}\,\left(\frac{r}{r_\mathrm{p}}\right)^{-3/7}\,\text{K} \quad\propto \frac{h^2}{r} \quad\Rightarrow\quad h(r) = h_\mathrm{p}\,\left(\frac{r}{r_\mathrm{p}}\right)^{2/7}
        \end{equation}
        for our irradiation prescription. By replacing the geometrical factor $\left(\D{\log H}{\log r}-1\right)h$ with a constant irradiation angle $\phi$ (e.g.,~\citet{dullemond-etal-2018}), we recover the familiar formula where $T\propto r^{-1/2}$ and $h\propto r^{1/4}$ instead (not used here).
                
        After constructing our initial surface density and temperature profiles, we first ran a one-dimensional radiative simulation without a planet and verified that the profiles indeed correspond to an equilibrium state. We then embedded the planet, and using the power-law profiles given in Eqs.~\eqref{eq:initial-surface-density-HD},~\eqref{eq:initial-surface-density-AS},~and~\eqref{eq:initial-temperature} as initial conditions, executed various radiative as well as locally isothermal simulations.
        
        In all setups, $\Sigma$ is damped to the initial profiles at the boundaries according to \citet{devalborro-etal-2006} over a timescale of 0.3 boundary orbital periods. The planet does not accrete material or migrate through the disk, and traces circular orbits around the star. The target simulation time was 1000~orbits for HD~163296---which at 48\,au translates into roughly 230\,kyr--- and 5000~orbits for AS~209 (5.4\,Myr at 99\,au) because these models have a very low viscosity of $10^{-5}$ . 
        We used a resolution of $573\times1118$ and $673\times1312$ cells for HD~163296 and AS~209, respectively, which results in square cells with 10-12 cells per scale height in the radial direction at the planet's location.
        
        Similar to \citet{ziampras-etal-2020}, the planet's presence is described by an additional gravitational force. We accounted for the shift of the system barycenter that is caused by the planet, but neglected backreaction of the disk onto the star and planet.
        \begin{table*}[h]
        	\centering
        	\caption{Physical and numerical parameters used in our modeling of the sources HD~163296 and AS~209. The opacity model by \citet{lin-papaloizou-1985} (here referred to as LP85) functionally translates to $\kappa=5\times 10^{-4}\,T^2$ within our simulation domain ($T<170$\,K). Only the two models labeled ``base'' (columns 1~and~7) are discussed in the results section; the remaining models (columns 2--7) were used to verify our results and are discussed in Appendix~\ref{appendix:locally-iso-limit}. Dashes imply that a parameter is inherited from the base model (column 1 for HD~163296). In our radiative models, the energy equation allows temperature-related quantities (i.e., $h_\mathrm{p}$, $T_\mathrm{p}$, $\kappa$ if defined by LP85) to evolve throughout the simulation. These are otherwise kept fixed in the corresponding locally isothermal runs.}
        	\begin{tabular}{c| c c c c c c | c}
        		\hline
        		system & \multicolumn{6}{c|}{HD~163296} & AS~209\\\hline
        		parameter & (1) base & (2) $\gamma\approx 1$ & (3) low $\kappa$ & (4) $\kappa$ by LP85 & (5) low $\alpha$ & (6) shallow $\Sigma$ & (7) base\\\hline\hline
        		$M_\ast$ [$\mathrm{M_\odot}$] 				& 2.089 & \filler 	& \filler 	& \filler 	& \filler 	& \filler 	& 0.83\\
        		$L_\ast$ [$\mathrm{L_\odot}$] 				& 16.98 & \filler 	& \filler 	& \filler 	& \filler 	& \filler 	& 1.41\\
        		$M_\mathrm{p}$ [$\mathrm{M_J}$] 			& 0.5 	& \filler 	& \filler 	& \filler 	& \filler 	& \filler 	& 0.83\\
        		$M_\mathrm{p}$ [$10^{-3}\,M_\ast$] 			& 0.29 	& \filler 	& \filler 	& \filler 	& \filler 	& \filler 	& 0.1\\
        		$M_\mathrm{disk}$ [$\%\,\mathrm{M}_\odot$] 	& 5.58	& \filler 	& \filler 	& \filler 	& \filler 	& 7.37  	& 68.1\\
        		$M_\mathrm{disk}$ [$\%\,M_\ast$] 			& 2.67 	& \filler 	& \filler 	& \filler 	& \filler 	& 3.53 		& 82.0\\
        		$r_\mathrm{p}$ [au] 						& 48 	& \filler 	& \filler 	& \filler 	& \filler 	& \filler 	& 99\\
        		$\Sigma_{t=0}(r)$ 			& $\propto r^{-3/2}$ 	& \filler 	& \filler 	& \filler 	& \filler 	& $\propto r^{-1}$ & $\propto r^{-1}$\\
        		$\Sigma_\mathrm{p}$ [g/cm${}^2$] 			& 9.61 	& \filler 	& \filler 	& \filler 	& \filler 	& \filler 	& 10.0\\
        		$h_\mathrm{p}$ [\%] 						& 5.601 & \filler 	& \filler 	& \filler 	& \filler 	& \filler	& 8.175\\
        		$T_\mathrm{p}$ [K] 							& 34.5	& \filler 	& \filler 	& \filler 	& \filler 	& \filler	& 14.2\\
        		$\gamma$ 									& 1.4 	& 1.01	 	& \filler 	& \filler 	& \filler 	& \filler	& 1.4\\
        		$\kappa$ [cm${}^2$/g] 						& 0.45 	& \filler 	& 0.045 	& LP85	 	& \filler 	& \filler	& LP85\\
        		$\log\alpha$  								& $-4$ 	& \filler 	& \filler 	& \filler 	& $-5$	 	& \filler	& $-5$\\
        		
        		$r_\mathrm{min}$--$r_\mathrm{max}$ [$r_\mathrm{p}$]
        		& 0.2--5	& \filler 	& \filler 	& \filler 	& \filler 	& \filler	& 0.1--10\\
        		$r_\mathrm{min}$--$r_\mathrm{max}$ [au] & 9.6--240	& \filler 	& \filler 	& \filler 	& \filler 	& \filler	& 9.9--990\\
        		$N_r\times N_\phi$ 				& $573\times 1118$ 	& \filler 	& \filler	& \filler 	& \filler 	& \filler	& $673\times1312$\\
        		$N_\mathrm{cells}/H_\mathrm{p}$				& 10 	& \filler 	& \filler 	& \filler 	& \filler 	& \filler	& 12\\
        	\end{tabular}		
        	\label{table:HD-VS-AS}
        \end{table*}
		To prevent singularities near the planet, we introduced a softening length $\epsilon=0.6H$ following \citet{mueller-etal-2012}, using the local pressure scale height and a correction factor that accounts for the disk thickness.
        
        \newpage
        \section{Results}
        \label{section:results}
        
        In this section we present the results of our numerical simulations. As stated above, we tested several different models. We focus in this section on the locally isothermal and radiative models with parameters described in the previous section. The remaining models are used for comparison purposes and are discussed in Appendix~\ref{appendix:locally-iso-limit}.
        
        First, we present our results on HD~163296. We start with focusing on overall disk profiles and differences between the locally isothermal and radiative models in terms of the radial distribution of gas in the disk. We then compare and contrast the structure of spiral arms between the two models in terms of surface density and temperature contrast with respect to the disk background. This is then followed up by results on AS~209 regarding the first point only because this is the most crucial to the discussion.
        
        \subsection{HD~163296}
        
        The first system we studied is HD~163296. This system features several rings, as shown in Fig.~\ref{fig:dsharp}.
        \subsubsection{Disk profiles and secondary gaps}
        \label{section:results-profiles}
        
        \begin{figure}
                \includegraphics[width=.5\textwidth]{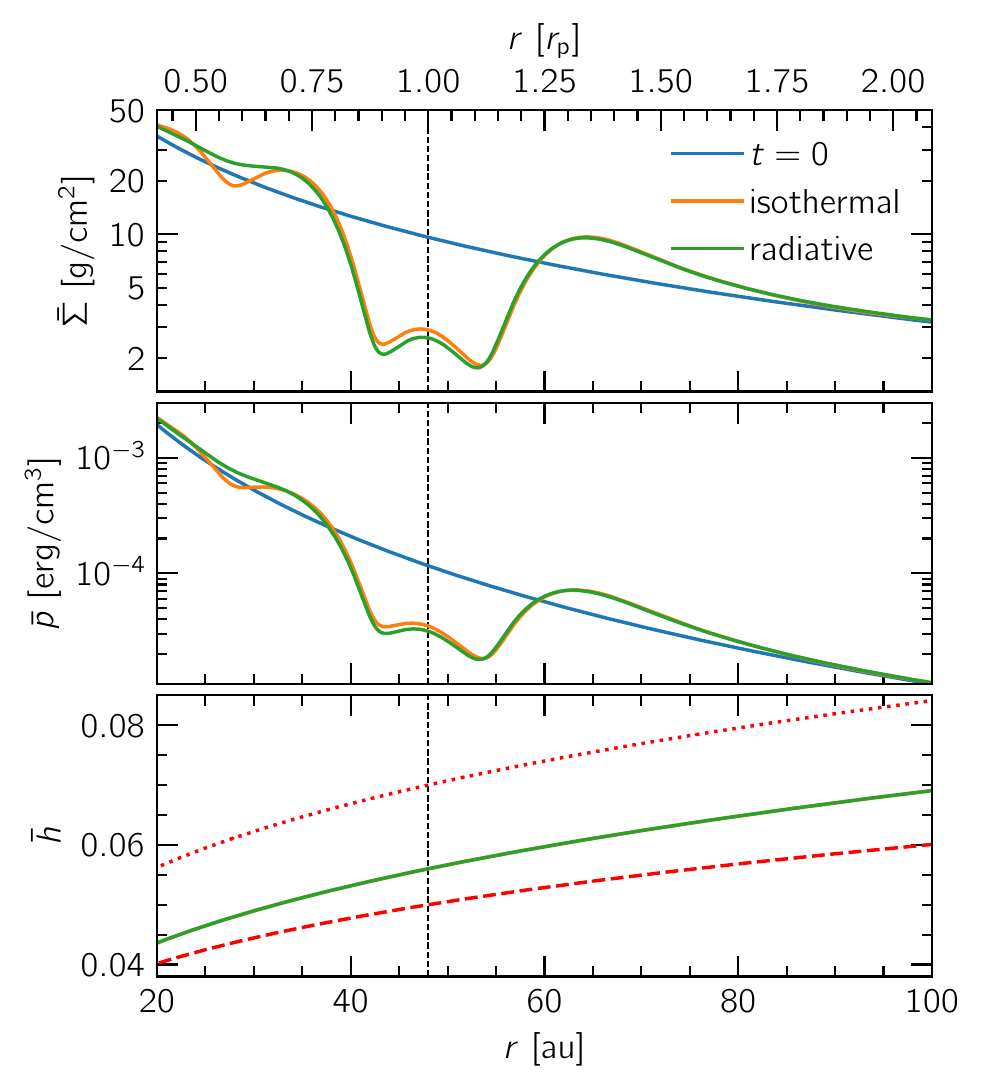}
                \caption{Azimuthally averaged surface density, midplane pressure, and aspect ratio (used as a proxy for temperature, see eq.~(\ref{eq:initial-temperature})) of HD~163296 for a locally isothermal (in orange) and a radiative (in green) disk after 1000 orbits with an embedded planet with mass $0.5\,\mathrm{M_J}$ at 48\,au. 
                Vertical black lines mark the location of the planet. The dashed and dotted red lines in the bottom panel denote the aspect ratio profiles used in the simulations by \citet{zhang-etal-2018}, where $h\propto r^{1/4}$ and $h_\mathrm{p}=0.05$ and 0.07, respectively. The fact that our aspect ratio profile is bounded by these two curves allows us to validate our results by comparing them to that study (see Section~\ref{section:discussion}).}
                \label{fig:isorad}
        \end{figure}

        Our main results are plotted in Fig.~\ref{fig:isorad}. By comparing our radiative and isothermal runs and computing the aspect ratio as a proxy for temperature (because $h\propto\sqrt{rT}$), we find that the azimuthally averaged temperature profiles of the two are practically identical with each other (bottom panel in Fig.~\ref{fig:isorad}), and therefore with the profile that an irradiation-dominated disk should have.
                \begin{figure}
                        \includegraphics[width=.5\textwidth]{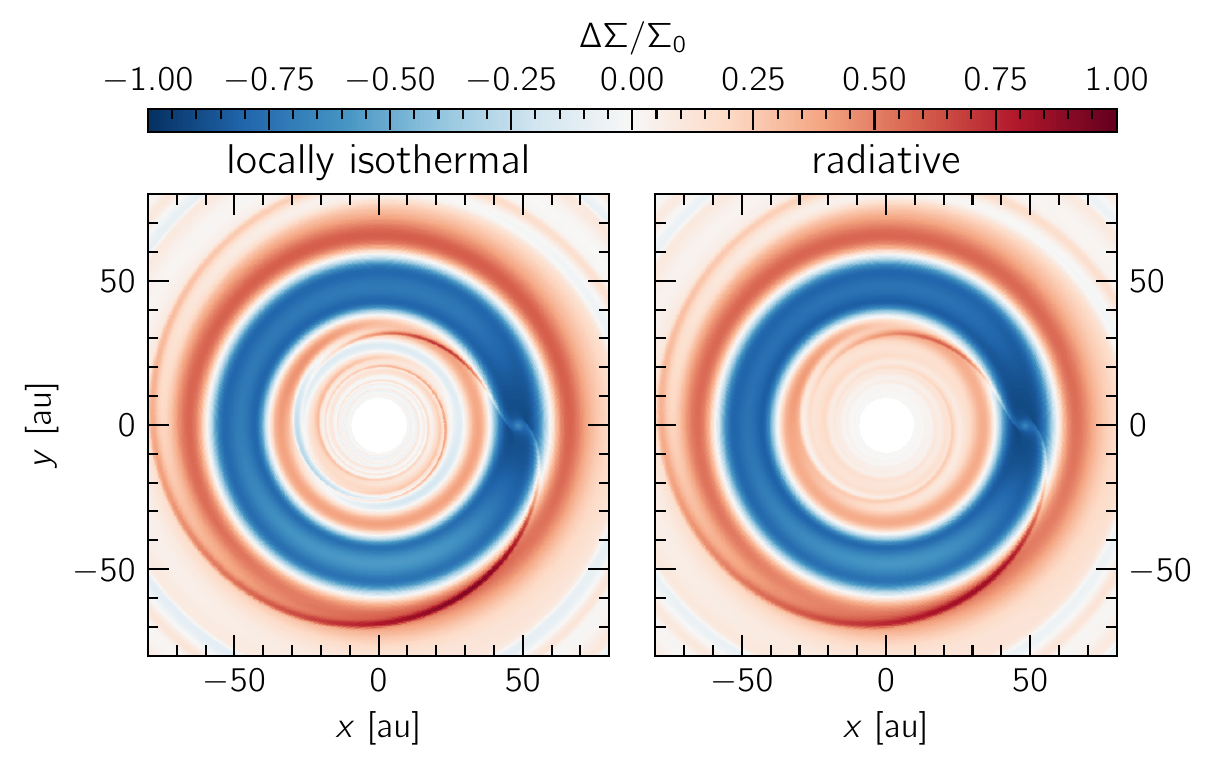}
                        \caption{Surface density perturbation with respect to the initial profiles for the two models of HD~163296. The locally isothermal run features a secondary gap at roughly 28\,au and slightly more prominent inner spirals.}
                        \label{fig:density-map}
                \end{figure}
        However, the azimuthally averaged surface density profiles of the two models differ in the inner disk and around the planet within 1000 orbits. The difference lies in the additional depression of the gas surface density at around 28\,au for the locally isothermal case, in contrast to the smoother profile in the radiative run. This secondary gap, while shallow, can warrant the existence of a pressure bump in the inner disk, creating a dust trap and therefore a second and potentially visible dust ring. 
        In contrast to the isothermal run, this pressure minimum disappears in the radiative simulation (middle panel in Fig.~\ref{fig:isorad}).
        
        A rather small (compared to that at 28\,au) but visible difference can also be seen within the gap region. This may be related to nonlinear effects caused by the planet's gap opening, and discuss this in Section~\ref{section:discussion}. Finally, the azimuthally averaged surface density in the outer disk is indistinguishable between the two models.
        
        In Fig.~\ref{fig:density-map} we compare the surface density in the disk to its initial profile. A gap and ring are clearly visible at 28 and 34\,au, respectively, in the locally isothermal model. Additionally, spiral arms in the inner disk have slightly higher contrast with the background for that model compared to its radiative counterpart. We carry out a more detailed comparison in the next section.
        
        \subsubsection{Spiral arm contrast}
        \label{section:results-spirals}
        
        As mentioned in the work of \citet{miranda-rafikov-2019a}, the locally isothermal equation of state overestimates the contrast of structures in protoplanetary disks. We already saw this behavior in Fig.~\ref{fig:isorad}, where a secondary gap is visible at 28\,au. However, nonaxisymmetric features in the disk such as spiral arms are lost when averaging along annuli. With this in mind, it is useful to track the spiral arms that are excited by the planet and compare the surface density along their crests with its azimuthally averaged values. In doing so, we can estimate their contrast with respect to the disk background.

        \begin{figure}[h]
                \includegraphics[width=.5\textwidth]{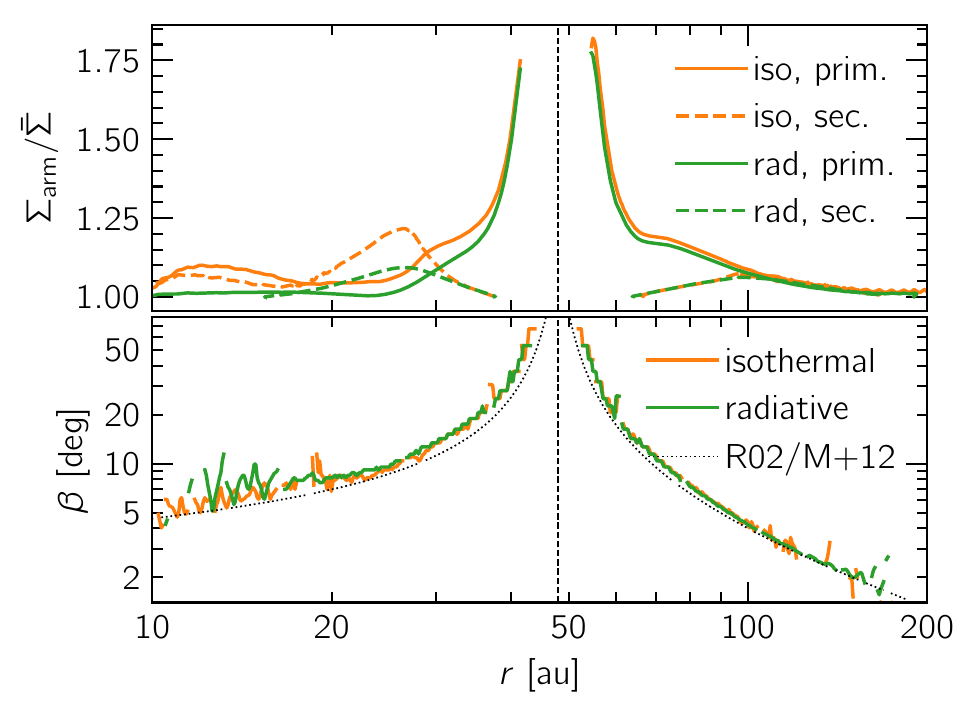}
                \caption{Comparison of spiral arm properties between our two models for HD~163296. Top: Primary and secondary spiral arm contrast with respect to the disk background. We find that the locally isothermal model shows consistently higher contrast. 
                        Bottom: Pitch angle of spirals as a function of radius. Because the two models have identical aspect ratios, the overlap of the two curves is to be expected. The vertical black line marks the planet location. The dotted black line corresponds to the analytical formulae in \citet{rafikov-2002}, Eq.~(44), and \citet{muto-etal-2012}, Eq.~(1).}
                \label{fig:spirals}
        \end{figure}

\begin{figure}
        \includegraphics[width=.5\textwidth]{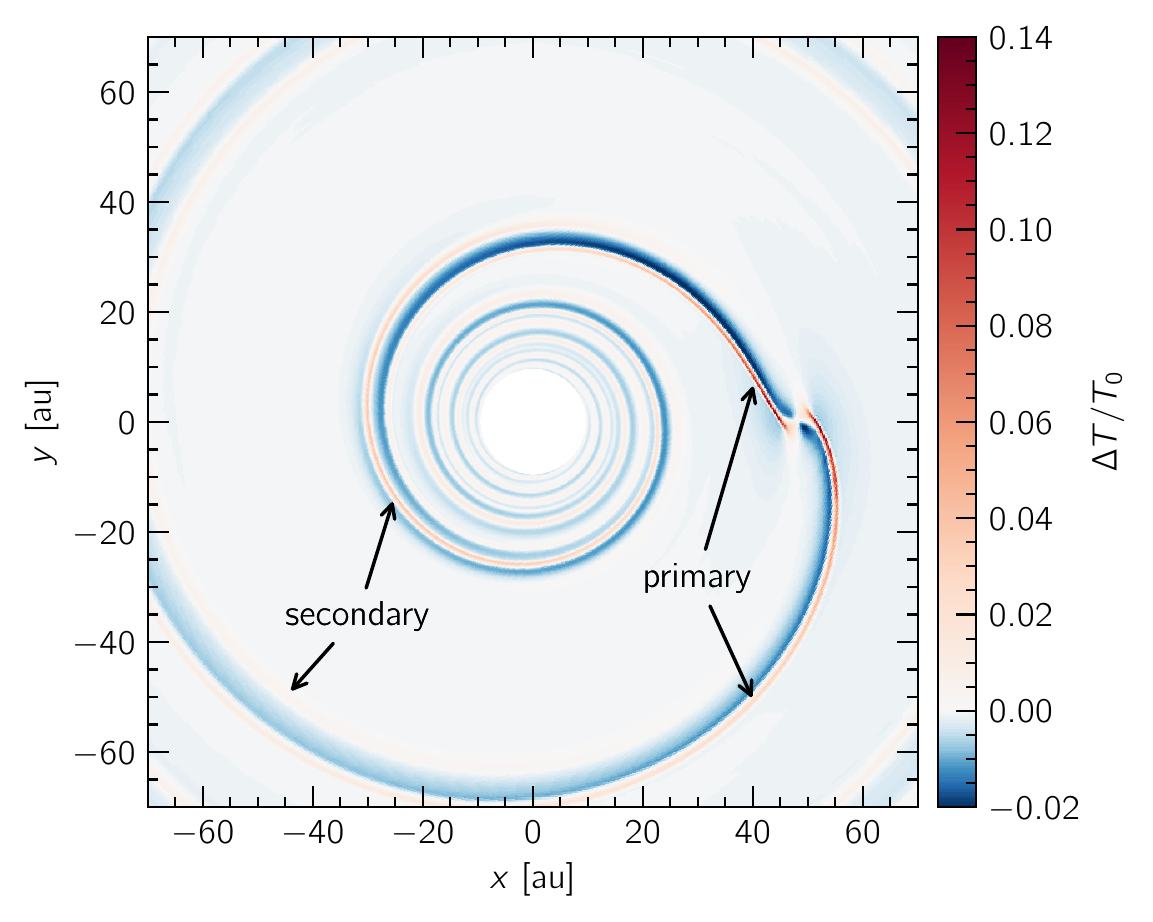}
        \caption{Temperature contrast of spirals with respect to the disk background for the radiative run of HD~163296. The effect of spiral heating on the global disk is negligible, and the temperature along spiral crests can increase by up to 15\%.}
        \label{fig:heatmap}
\end{figure}

        This comparison was carried out by tracking the trajectory of each individual spiral, logging the surface density $\Sigma_\mathrm{arm}$ along their peaks and then plotting the ratio $\Sigma_\mathrm{arm}/\bar{\Sigma}$, where $\bar{\Sigma}$ is the azimuthally averaged surface density (see \citet{ziampras-etal-2020} for details). Our results are summarized in Fig.~\ref{fig:spirals} and show that spirals in the locally isothermal model are indeed consistently "stronger" (i.e.,~have a higher contrast with respect to the disk), with the effect being more prominent in the inner disk (top panel).
        
        The pitch (or opening) angle $\beta$ of these spirals can be defined (e.g., \citet{zhu-etal-2015}) through
        \begin{equation}
                \label{eq:pitch-angle}
                \tan\beta\equiv\frac{\mathrm{d}r}{r\mathrm{d}\phi} \approxeq \frac{c_\mathrm{s}}{r|\Omega(r)-\Omega_\mathrm{p}|} = \frac{h}{|1-(r/r_\mathrm{p})^{3/2}|}.
        \end{equation}
        Equation~\eqref{eq:pitch-angle} suggests that the pitch angle of spirals should be a function of the aspect ratio (again, a proxy for the temperature) for a given distance $r$. With this in mind, and given that the two models share an identical radial temperature profile, we expect and observe that the pitch-angle profiles of both primary and secondary spirals match very well for the locally isothermal and radiative model (lower panel).

        We also compared the temperature along spiral crests with the initial axisymmetric profile for the radiative model. We found that differences are about 10-15\% or smaller, and that spiral heating on the entire disk is negligible (see Fig.~\ref{fig:heatmap}). This is to be expected in the optically thin irradiation-dominated region of the disk at $r\gtrsim20$\,au (e.g., \citet{rafikov-2016}).
        
        \subsection{AS~209}
        \label{section:results-as209}
        
        \begin{figure}
                \includegraphics[width=.5\textwidth]{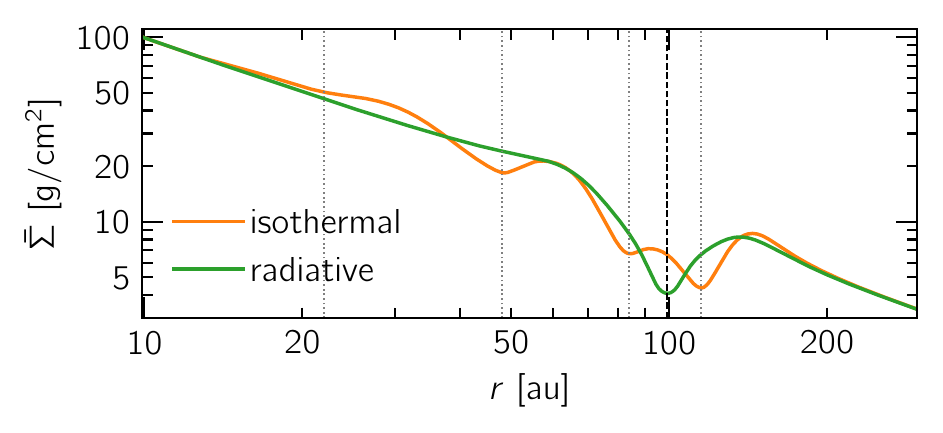}
                \caption{Surface density profile for our numerical model of AS~209 for a locally isothermal and a radiative equation of state after 5000 orbits. The differences in the inner disk as well as the gap region are clear between the two models. Vertical dotted lines mark location candidates for dust gaps based on the formula determined by \citet{zhang-etal-2018}, and the dashed line marks the planet location.}
                \label{fig:AS209}
        \end{figure}
        
        \begin{figure}
                \includegraphics[width=.5\textwidth]{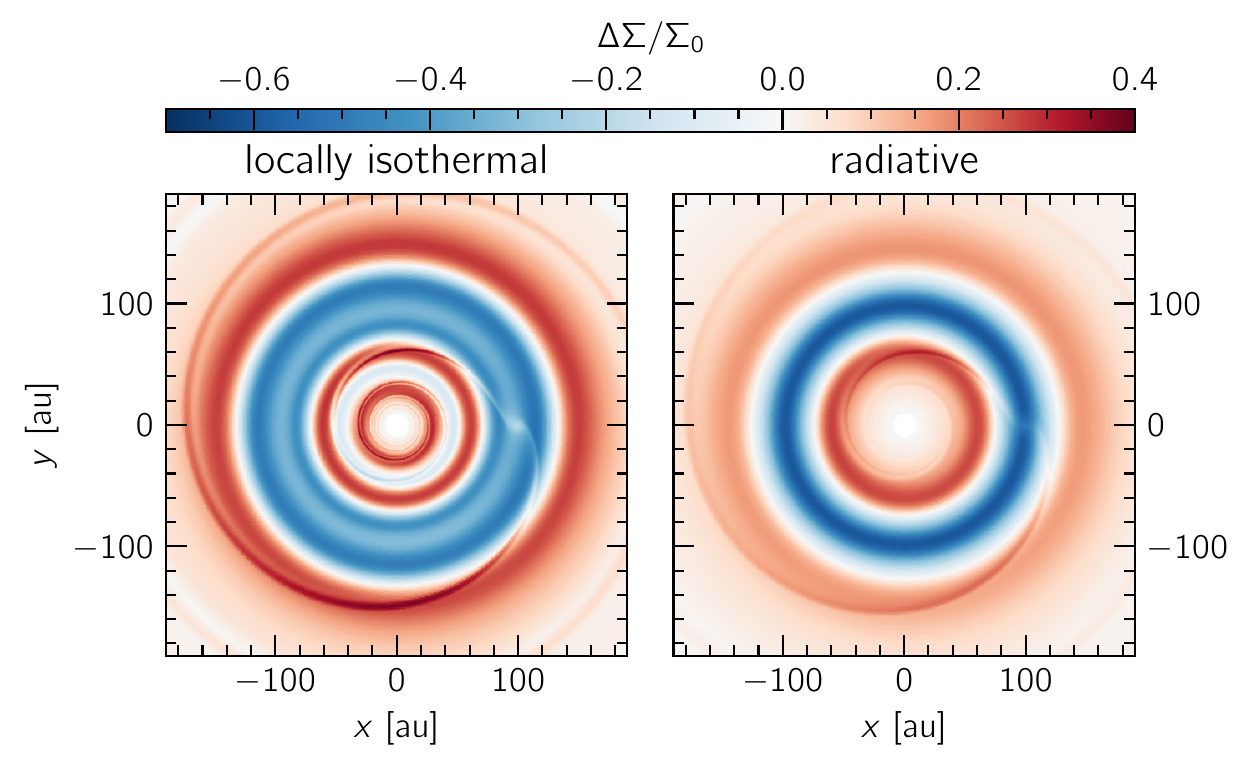}
                \caption{Similar to Fig.~\ref{fig:density-map}, but for AS~209. The sharper annular and angular structures in the locally isothermal model are clearly visible.}
                \label{fig:density-map-as209}
        \end{figure}
        

        The goal in modeling this additional source was to see how this disparity between the locally isothermal and adiabatic equations of state would affect the gas surface density profile of a system with more gaps observable in continuum emission: five in AS~209 (at $r\in\{24, 35, 61, 90, 105\}$\,au) compared to only one in HD~163296 (at $r=48$\,au) for our simulation domain \citep{huang-etal-2018}. While reproducing all five gaps might be difficult with an aspect ratio $h_\mathrm{p}\approx0.08$, we see substantial differences between the two setups not only in the number of producible gaps, but also in the shape of the primary gap around the planetary orbit.
        
        Our results are plotted in Figs.~\ref{fig:AS209}~and~\ref{fig:density-map-as209} and match our findings for HD~163296 in that the radiative model does not agree with the locally isothermal one. A single planet at 99\,au reproduces at least three gaps at 47, 84 and 115\,au in the locally isothermal model (with a possible fourth gap at 22\,au, which is however subject to boundary effects), whereas the radiative model produces a very smooth monotonic profile with no gaps except for one exactly at the planet location (Fig.~\ref{fig:AS209}). In addition, the primary gap structure is entirely different between the two models, with the locally isothermal model showing a wider gap region that contains more material at the orbital radius of the planet, however. This can be seen in both figures, and is discussed in detail in Section~\ref{section:discussion}.

        \section{Discussion}
        \label{section:discussion}

        In Section~\ref{section:results} we showed that modeling a protoplanetary disk with a locally isothermal equation of state can under certain circumstances prove incorrect because this assumption might affect the planet--disk interaction process even when the temperature profile is largely unaffected by the choice of equation of state. By comparing against a radiative simulation where viscous and irradiation heating and thermal cooling are included, we showed that radiative effects play an important role in the disk evolution even at large distances from the source star. 
        
        Our testbeds aimed at reproducing the inner 50 au structure of the system HD~163296 and the gap-rich structure of AS~209. At distances of 48\,au for HD~163296 and 99\,au for AS~209 (which is where planets are suspected to be), the disk is strongly dominated by irradiation, whereas viscous and spiral heating effects are negligible. Even so, the cooling timescale at the range of 20--100\,au in these systems is still significant enough to invalidate the locally isothermal assumption.
        
\begin{figure}[h]
                \includegraphics[width=.5\textwidth]{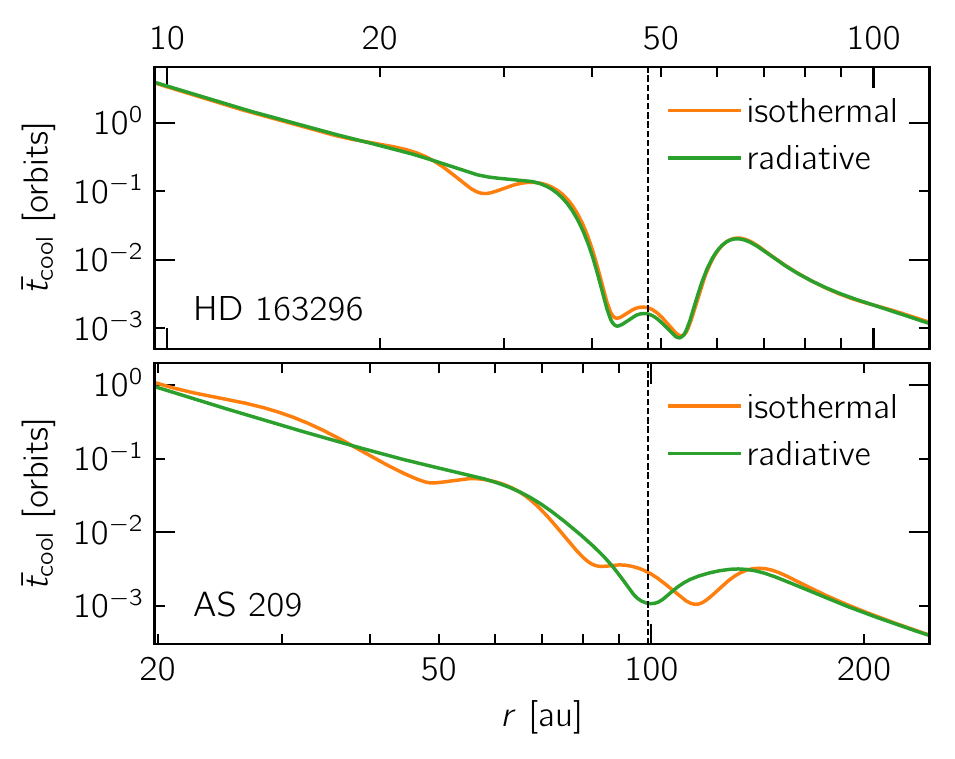}
                \caption{Azimuthally averaged cooling timescale in units of the local orbital period of the disk after 1000 and 5000 planetary orbits for HD~163296 and AS~209, respectively. Differences between locally isothermal and radiative models become clear in the inner disk, where the cooling timescale corresponds to a non-negligible fraction of the orbital period (roughly 1--10\% near the planet, and comparable to the orbital period farther in). In the outer disk, cooling is sufficiently fast so that the two models overlap as far as azimuthally averaged profiles are concerned.}
                \label{fig:tcool}
        \end{figure}

        We can estimate the cooling timescale as
        \begin{equation}
        \label{eq:cooling-timescale}
        \frac{\partial\Sigma e}{\partial t}\sim \frac{\Sigma e}{t_\mathrm{cool}} \sim Q_\mathrm{cool} \Rightarrow t_\mathrm{cool} \propto \frac{\tau_\mathrm{eff}\Sigma }{T^3}.
        \end{equation}
        We then compare $t_\mathrm{cool}$ to the local orbital period $P_\mathrm{orb}=2\pi/\Omega_\mathrm{K}$ in Fig.~\ref{fig:tcool}. We find that $t_\mathrm{cool}$ corresponds to 2--20\% of $P_\mathrm{orb}$ in the 20--50\,au range in the absence of a planet for HD~163296, and 1--10\% in the 40--100\,au range for AS~209. These fractions are clearly significant in this context, implying that the angular momentum flux driven by the spiral waves of a planet is described by an adiabatic framework \citep{miranda-rafikov-2019a}.
        
        At the same time, we see that the cooling timescale drops below 0.1\% of the orbital period at distances of $\sim$100\,au for HD~163296, or $\sim$200\,au for AS~209 (see Fig.~\ref{fig:tcool}). At these distances we find that the locally isothermal model approaches the radiative one both in terms of radial disk structure and spiral arm contrast, suggesting that the locally isothermal assumption can still be justified when the cooling timescale is short enough. This is consistent with the latest results by \citet{miranda-rafikov-2019c}, who suggested that a cooling timescale of about $10^{-3}$--$10^{-2}$ orbital periods is sufficiently short to match locally isothermal disks for low- and high-mass planets, respectively. Regarding this last statement, the different gap structure between equations of state around the planetary orbit in AS~209 can be explained by comparing the planet mass to the thermal mass of the disk \citep{zhu-etal-2015}
        \begin{equation}
                \label{eq:thermal-mass}
                M_\mathrm{th}\equiv \frac{c_\mathrm{s}^3}{\mathrm{G}\Omega_\mathrm{p}}=h_\mathrm{p}^3 M_\ast \approx 1\,\mathrm{M_J}\, \left(\frac{h_\mathrm{p}}{0.1}\right)^3\left(\frac{M_\ast}{\mathrm{M}_\odot}\right).
        \end{equation}
        Using this formula, we find that a 0.5\,$\mathrm{M_J}$ planet in HD~163296 corresponds to 1.3\,$M_\mathrm{th}$, and a 0.083\,$\mathrm{M_J}$ planet in AS~209 to 0.18\,$M_\mathrm{th}$. This means that the former system is prone to nonlinear effects due to gap opening, and therefore the surface density profiles around the primary gap are largely the same between the two different equations of state. On the other hand, our numerical model of AS~209 hosts a planet that falls in the linear regime, and therefore a cooling time of about 1\% of the orbital period is not short enough to allow the radiative and locally isothermal models to agree with each other. Instead, the two profiles show a substantially different structure in the region that corotates with the planet, and they only overlap at the outskirts of the disk (where planet-driven effects are weak and/or damped anyway).

        To verify that the locally isothermal profile can indeed be recovered with a more efficient cooling prescription, we executed several additional radiative simulations of HD~163296 where we limited adiabatic effects ($\gamma=1.01$) or artificially reduced the cooling timescale (using $\kappa=0.045$~$\mathrm{cm}^2/\mathrm{g}$), thereby limiting radiative effects. These results are shown in Appendix~\ref{appendix:locally-iso-limit}.
                
        While this exercise shows that the inner disk is prone to radiative effects in our simulations, this result should be taken with a grain of salt when comparing to observations. Both the optical depth and surface density (which depend on the opacity and initial conditions) are uncertain and not set in stone. Additionally, if a substantial part of the dust grains has grown into larger grains, the opacity (and in turn, the optical depth) could be much lower, reducing the cooling timescale.
        
        Our results on HD~163296 can also be compared to the 2D locally isothermal simulations by \citet{zhang-etal-2018}. In Fig.~2 of their study, they show that a planet with $M_\mathrm{p}=0.3~\mathrm{M_J}$ ($q=2.9\times10^{-4}$, in our case $q=2.3\times10^{-4}$) can open a secondary gap at roughly 0.6\,$r_\mathrm{p}$ (28\,au for our single-planet model of HD~163296) when the aspect ratio $h_\mathrm{p}$ is between 0.05--0.07 and scales with $r^{1/4}$. Because our computed aspect ratio lies within the profiles generated using those values (see Fig.~\ref{fig:isorad}), we expect exactly (or at least) one secondary gap in our locally isothermal simulation. The fact that we indeed observe it suggests that our results agree with \citet{zhang-etal-2018} in the locally isothermal limit. In addition, we observe the secondary gap at the location inferred by the fitting formula suggested in that study. However, we used a steeper surface density profile than they did ($\Sigma_0\propto r^{-3/2}$ as opposed to $r^{-1}$). We carried out two more simulations with a shallower surface density profile and found similar results (see Appendix~\ref{appendix:locally-iso-limit}).
        
        By taking into account the above points, we see that the secondary-gap-opening capabilities of a single planet are exaggerated in a locally isothermal disk, such that a secondary gap in gas surface density can form within that framework but not always when radiative effects are properly treated. In the case of HD~163296 this could constrain the time of planet formation because a ring at 34\,au is indeed not visible \citep{huang-etal-2018}. This assumes that our estimates of $\alpha$ among other model parameters are viable.                 
        
        \begin{figure}[h]
                \includegraphics[width=.5\textwidth]{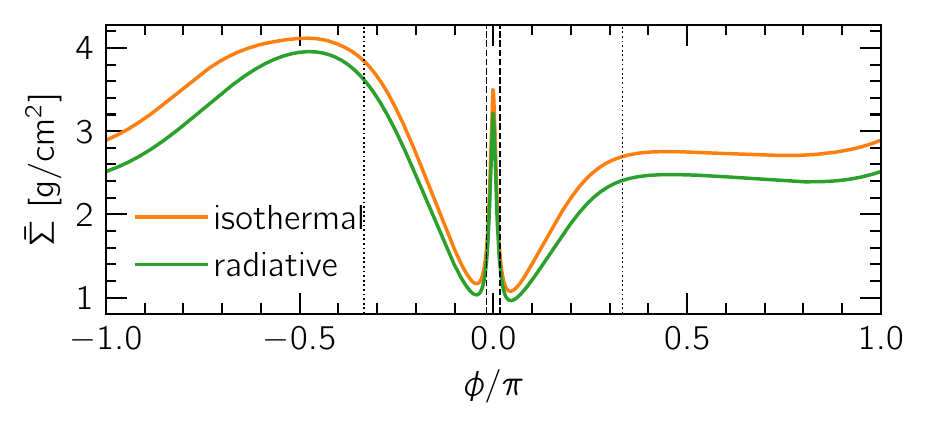}
                \caption{Azimuthal cut at $r=r_\mathrm{p}=48\,\mathrm{au}$. The surface density trailing behind the planet (left side, $\phi <0$) is slightly higher than that at the leading Lagrange point ($\phi >0$). The dashed black lines mark the planet's hill radius and contain 20 cells. The dotted lines mark the location of the L4 (left) and L5 (right) Lagrange points.}
                \label{fig:phiplot}
        \end{figure}

        Our simulations do not include a dust component, meaning that we cannot verify whether this secondary gap generated in the locally isothermal simulation is visible. Nevertheless, we expect that large grains can be trapped in the pressure bump formed at 34\,au, forming a ring and therefore increasing the contrast between the inner disk and the secondary gap region. A similar argument can be used for the crescent structure within the gap that can be seen in the ALMA observations of HD~163296. We find that the perturbed gas surface density at the trailing Lagrange point is higher than that at the leading one (see Fig.~\ref{fig:phiplot}), and we suggest that dust--gas interaction could collect grains at the L4 point. This feature, however, should be transient as the gap region continues to empty as the disk evolves.
        
        The system HD~163296 shows more structure between 50--200\,au, namely a 
        clear second gap in continuum emission at 86\,au and a third, slightly less visible gap at 145\,au. Assuming the existence of planets at all three locations, we can justify our results as far as annular structures in the 10--50\,au range are concerned because the planet at 48\,au will shield the inner disk by opening a gap and therefore halt the propagation of spiral waves by the outer planet(s). 
        
        Nevertheless, it would be interesting to include more planets and model the 10--200\,au range of this system, assuming that each gap in dust continuum stems from a single planet opening a corresponding gap in gas density. This is further motivated by the kinematic detection of Jupiter-sized planets at 83 and 137\,au by \citet{teague-etal-2018}. We therefore carried out a simulation with three planets at 48, 86, and 145\,au, respectively, and find similar results in the 10--50\,au range. These results are shown and discussed in Appendix~\ref{appendix:HD163296-multiplanet}.
        
        There are several other sources for which ALMA has provided high-fidelity observational datasets. As far as ring structures are concerned, one system stands out: AS~209 \citep{andrews-etal-2018} shows at least five rings with as many gaps in dust continuum emission \citep{huang-etal-2018}. Numerical models strongly suggest that one or more growing planets are responsible for their formation.
        
        It has been suggested that a single planet at 99\,au might be able to carve most of these gaps \citep{zhang-etal-2018}. We therefore found it useful to examine the importance of radiative effects on a system with such a rich annular structure, and carried out a comparison similar to that for HD~163296 against a locally isothermal simulation, shown in Section~\ref{section:results-as209}. Strikingly, while our locally isothermal model produces several rings throughout the disk, we find that this is not the case in the radiative model, which shows no ring structure except for the unavoidable pressure bump in the outer disk that is caused by the clearing of the planet's corotating region. Our results suggest that we would need multiple planets to explain the ring--gap structure in such systems.
        \section{Conclusions}
        \label{section:conclusions}
        
        Our goal was to examine the importance of radiative effects in disk evolution with regard to the gap-opening capabilities of a single planet. To this end, we carried out 2D numerical simulations of planets embedded in disks that resemble two systems that have recently been imaged in high angular resolution: HD~163296 and AS~209. We then compared locally isothermal simulations, where a fixed radial sound speed profile is prescribed, to setups where radiative effects are self-consistently treated.
        
        We found that locally isothermal models exaggerate the contrast of planet-generated features with respect to radiative models and therefore overestimate the planet's ability to carve a secondary gap in its disk. This is consistent with previous results and implies that a single planet cannot always explain the existence of multiple gaps.
        
        We also found that the contrast of spiral arms launched by a planet is artificially sharpened within a locally isothermal framework. While this phenomenon is weak or negligible in the optically thinner outer disk, it becomes more significant for inner spirals. Regardless, spiral shock heating is negligible, with spirals having a low temperature contrast with the background and the disk being sufficiently optically thin.
        
        Finally, by running a simulation of HD~163296 over an extended range that contained three planets, we found that our results in the limited range of 10--60\,au (i.e.,~inside and around the innermost planet) remain unchanged. We also showed that an interplay between planet mass and the cooling timescale can lead to slight differences between a locally isothermal and an adiabatic equation of state even at 145\,au, although viscous evolution might render such differences negligible.
                
        In conclusion, the locally isothermal assumption proves to be dangerous even at the range of tens of au regarding planet--disk interaction and should therefore be in general avoided in favor of an adiabatic equation of state with a prescription for radiative cooling in the disk. By estimating the cooling timescale $t_\mathrm{cool}$,  the usage of such an assumption in a regime where cooling occurs very rapidly compared to the orbital period $P_\mathrm{orb}$ might be justified. This corresponds to $t_\mathrm{cool}$ at least shorter than 1\% of $P_\mathrm{orb}$ for massive planets, or 0.1\% for low-mass planets, in good agreement with \citet{miranda-rafikov-2019c}.
        
        \begin{acknowledgements}
                
                CPD and WK acknowledge funding from the DFG research group FOR 2634 "Planet Formation Witnesses and Probes: Transition Disks" under grant DU 414/22-1 and KL 650/29-1, 650/30-1.
                The authors acknowledge support by the High Performance and
                Cloud Computing Group at the Zentrum f\"ur Datenverarbeitung of the University
                of T\"ubingen, the state of Baden-W\"urttemberg through bwHPC and the German
                Research Foundation (DFG) through grant no INST 37/935-1 FUGG.
                All plots in this paper were made with the Python library \texttt{matplotlib} \citep{hunter-2007}.
        \end{acknowledgements}
        
        \bibliographystyle{aa}
        \bibliography{refs}

        \begin{appendix}
                \section{Approaching the locally isothermal limit in HD~163296}
                
                \label{appendix:locally-iso-limit}

                To verify whether the locally isothermal results for HD~163296 concerning the primary and secondary gaps can be recovered by appropriately tweaking radiative effects, we tested several cases where we either restrained changes in internal energy by setting $\gamma=1.01$, or amplified the contribution of irradiation and cooling by lowering the opacity to $\kappa=0.045$\,$\mathrm{cm}^2/\mathrm{g}$. A detailed description of the model parameters used here is given in Table~\ref{table:HD-VS-AS}, columns 2 and 3.
                
\begin{figure}[h]
                        \includegraphics[width=0.5\textwidth]{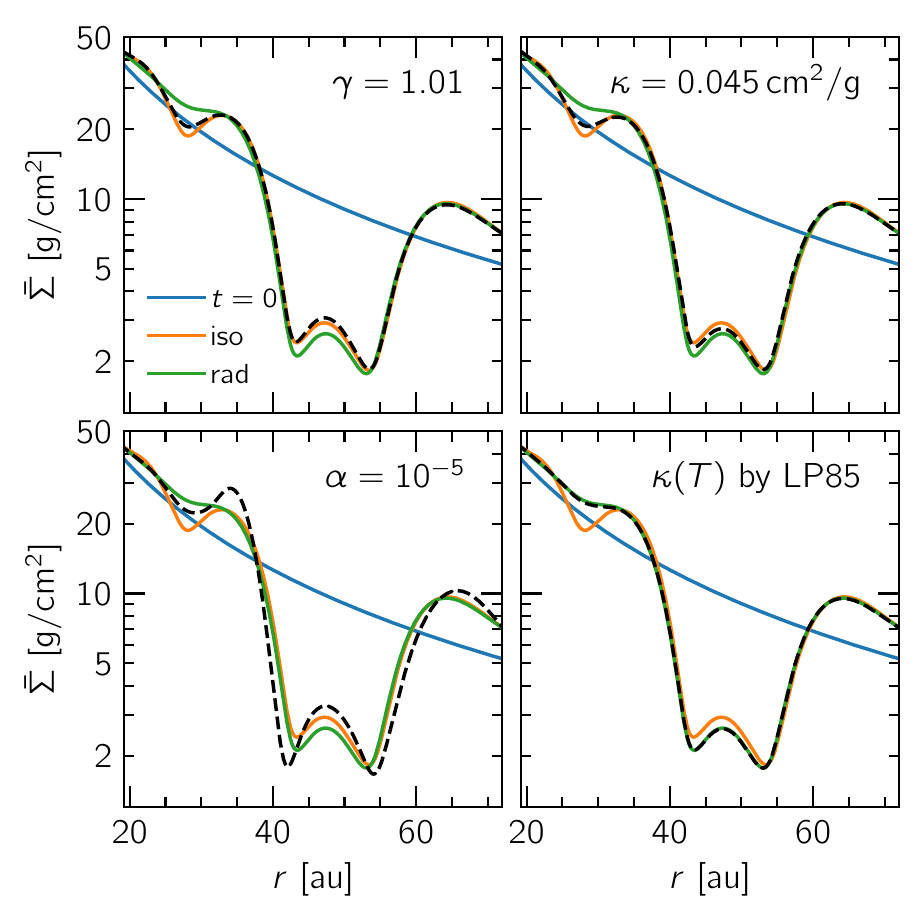}
                        \caption{Azimuthally averaged surface density profiles after 1000 orbits for several comparison tests, attempting to recover the locally isothermal limit by constraining radiative effects. The orange and green curves correspond to our fiducial locally isothermal and radiative models, and the change being tested in each panel is plotted with a black dashed line. An adiabatic equation of state with $\gamma=1.01$ or $\kappa=0.045$\,$\mathrm{cm}^2$/g (instead of 1.4 and 0.45\,$\mathrm{cm}^2$/g, respectively) roughly reconstructs the gap structure of the locally isothermal model and could allow the formation of a shallower secondary gap. A model with lower viscosity ($\alpha=10^{-5}$ instead of $10^{-4}$) also shows a secondary gap, but differences {in the width of the primary gap and in the outer disk} are also visible. The opacity model by \citet{lin-papaloizou-1985} yields no observable difference.}
                        \label{fig:isothermal-limit}
                \end{figure}

                Our results, plotted in Fig.~\ref{fig:isothermal-limit}, show that locally isothermal conditions can be emulated to an extent by manipulating the contribution of the energy equation altogether through the adiabatic index, or controlling the cooling timescale through the opacity (Eq.~\ref{eq:cooling-timescale} suggests that $t_\mathrm{cool}\propto\tau\propto\kappa$). It might be possible to completely recover the locally isothermal limit by further constraining $\gamma\rightarrow1$ or artificially lowering the opacity even further.
                
                In addition, to test the planet's secondary-gap opening capabilities under conditions that could potentially lead to the reemergence of the secondary gap, as well as to verify our opacity model of choice, we executed a few more pairs of locally isothermal and radiative simulations.
                
				We set $\alpha=10^{-5}$ in the first pair to facilitate the gap opening process. We find that this does lead to the surface density contrast (and the pressure bump) reappearing in the inner disk, but at the same time, it also results in a wider gap and a different outer disk profile (see Fig.~\ref{fig:isothermal-limit}). This value of $\alpha$ is extremely low and gap opening is known to be easier when the viscosity is low (e.g., \citet{crida-etal-2006}), therefore we argue that this adds nothing new to our results.
                        
                In the second pair, we compared our constant-opacity models and the analytical opacity model by \citet{lin-papaloizou-1985}, which dictates a relation $\kappa \propto T^2$ for $T<170$\,K (i.e.,~within our simulation domain). We find that the choice of opacity model has little to no effect as far as the secondary-gap opening capabilities of the planet are concerned (Fig.~\ref{fig:isothermal-limit}). This makes sense because irradiation heating and thermal cooling dominate the energy equation, and equating these two together factors the opacity out given our prescription in Eq.~\eqref{eq:source-terms} while still yielding a similar value for the cooling timescale at the 20--40\,au region.
                
\begin{figure}[h]
                        \includegraphics[width=0.5\textwidth]{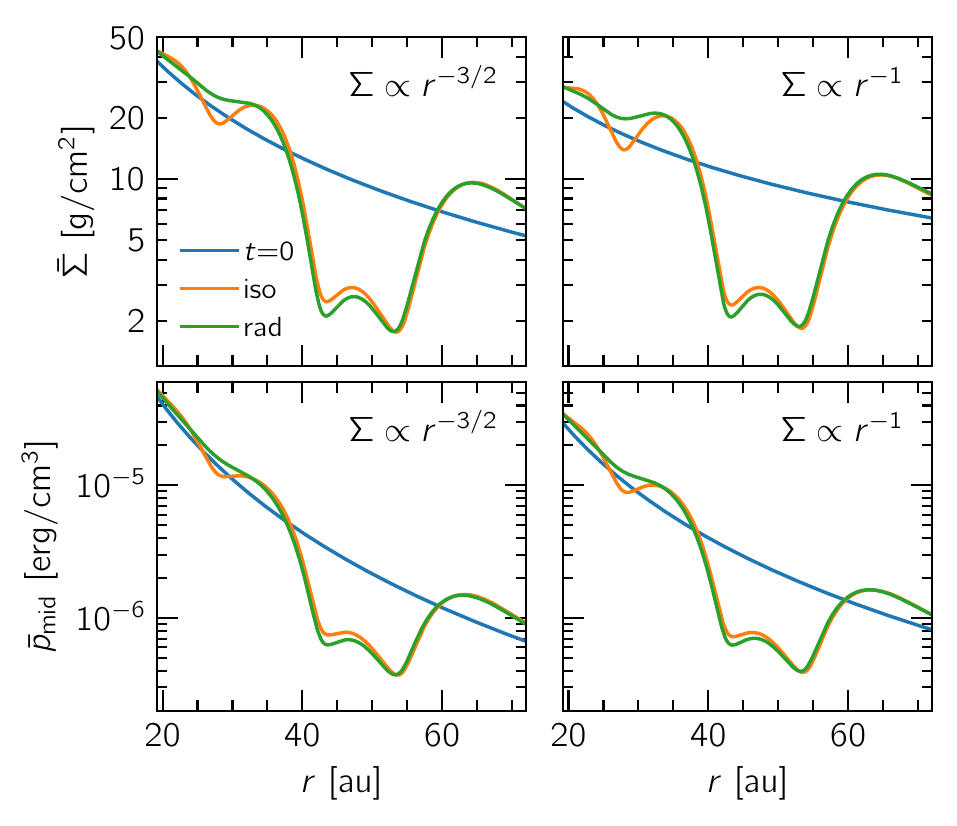}
                        \caption{Azimuthally averaged surface density and midplane pressure profiles after 1000 orbits for two comparison tests where disks with different surface density profiles are simulated. In the right panels, a shallower surface density profile ($\Sigma_0 \propto r^{-1}$ instead of $\propto r^{-3/2}$) is used.}
                        \label{fig:different-disks}
                \end{figure}

                Finally, in the third pair we prescribed a shallower initial surface density profile $\Sigma(r) \approx 10\,(r/r_\mathrm{p})^{-1}$ in an attempt to reduce the cooling timescale in the inner disk while preserving the conditions in the planet's vicinity. In doing so, we can also compare against the results from locally isothermal simulations by \citet{zhang-etal-2018}. The results are summarized in Fig.~\ref{fig:different-disks} and paint a picture similar to that for the steeper, $\Sigma(r) \propto r^{-3/2}$ profile. However, because the cooling timescale is shorter for the shallower $\Sigma$ profile (by a factor of 30\% and 40\% at the location of the secondary pressure bump and gap, respectively), the disk acts more "locally isothermally" and the secondary density bump is still visible after 1000 orbits but a pressure minimum does not form in the midplane. At the same time, pressure torques near the secondary gap edge are weaker for the shallower profile; this effect assists the gap-opening process.
                
				A detailed description of the model parameters used in these models is given in Table~\ref{table:HD-VS-AS}, columns 4, 5, and 6.
                \newpage
                \section{Modeling HD~163296 with three planets}
                \label{appendix:HD163296-multiplanet}
                
                As mentioned in Section~\ref{section:discussion}, we conducted an additional simulation of HD~163296 with three planets at 48, 86, and 145\,au to examine the structure of the system on a larger scale, more easily comparable to the DSHARP observation (see Fig.~\ref{fig:dsharp}, left panel). To do so, we employed a fourth-order Runge--Kutta n-body integrator \citep{thun-kley-2018} that allows the planets to interact gravitationally and accounts for the noninertial term that arises due to centering our system around $r=0$ instead of the barycenter of the system.
                        
                In our model, all three planets have the same mass of 0.5\,$\mathrm{M_J}$. Because the thermal mass of the disk scales purely with the aspect ratio at a planet's location, we expect a damping of nonlinear effects as we slowly transition to the linear regime upon moving farther out in the disk. More specifically, the ratio $M_\mathrm{p}/M_\mathrm{th}$ is 1.3, 0.8, and 0.5 for the three planets in ascending distance from the star. Because the cooling timescale at 86 and 145\,au is 0.33\% and 0.06\% of the local orbital period, respectively, it is possible that we can still observe deviations between the locally isothermal and radiative models around the planets' corotating regions because we now study the low-mass regime.
                        
\begin{figure}[h]
                        \includegraphics[width=0.5\textwidth]{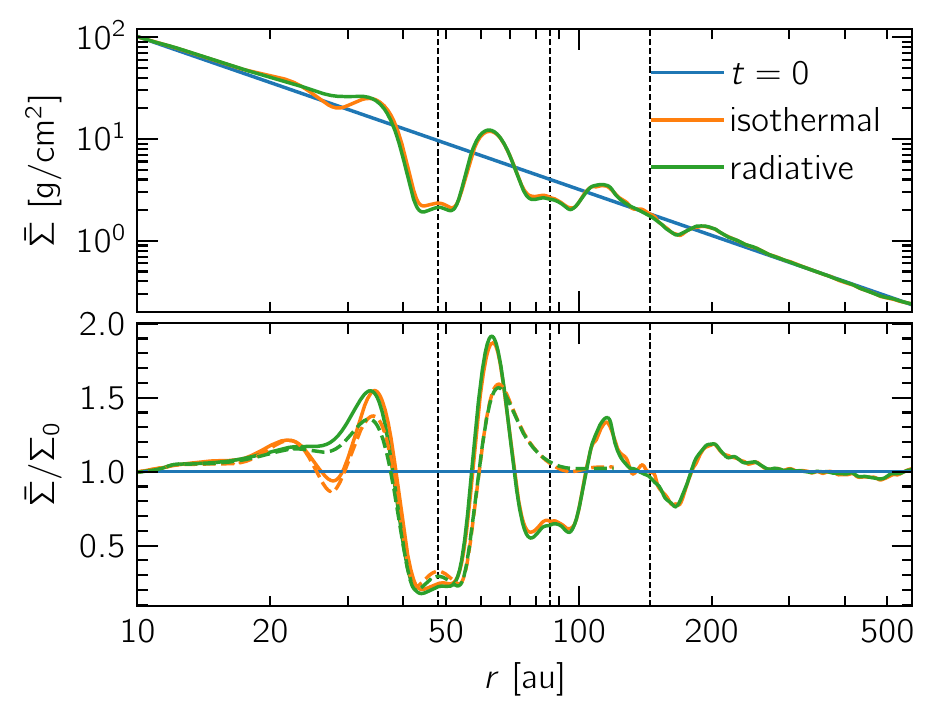}
                        \caption{Azimuthally averaged surface density profile after 1000 orbits at 48\,au (417 at 86\,au, 190 at 145\,au) for the three-planet model (see Table~\ref{table:HD-multiplanet}). The colored dashed lines refer to the base model with a single planet at 48\,au. Vertical black lines show the location of planets in the disk. }
                        \label{fig:isorad-3-planets}
                \end{figure}

                The resulting surface density profile is plotted in Fig.~\ref{fig:isorad-3-planets}. As expected, the two models show better overall overlap with increasing distances from the star. Nevertheless, small deviations are still visible within the corotating regions of the outer planets. We rationalize this outcome both through the cooling timescale argument and by considering how the interaction between a planet and the disturbances caused by other planets (e.g., spirals) could affect their corotating regions. It should be noted, however, that the viscous timescale at 145\,au is roughly three times longer than that at 48\,au, so it is possible that differences between equations of state can emerge on much longer timescales, which might not be reasonable for such a young system.
                
                Finally, by comparing the 10--60\,au range of the three-planet simulation against that of our base model with a single planet (lower panel of Fig.~\ref{fig:isorad-3-planets}) we find that the multi-planet system shows stronger perturbations in surface density around the secondary gap at 28\,au, but still fails to form a pressure maximum at 34\,au in the radiative model. This helps support the relevance of our base model, which targeted a limited range of HD~163296.
                
                %
                        \begin{table}[h]
                        \centering
                        \caption{Parameters used for the model of HD~163296 containing three planets, compared to the base model of the same system. Dashes imply that a parameter is inherited from the base model.}
                        \begin{tabular}{c| c c }
                                \hline
                                parameter & base & three planets\\\hline\hline
                                $M_\ast$ [$\mathrm{M_\odot}$]                           & 2.089 & \filler \\
                                $L_\ast$ [$\mathrm{L_\odot}$]                           & 16.98 & \filler \\
                                $M_\mathrm{p}$ [$\mathrm{M_J}$]                         & 0.5     & 0.5 / 0.5 / 0.5 \\
                                $M_\mathrm{p}$ [$10^{-3}\,M_\ast$]                      & 0.29    & 0.29 / 0.29 / 0.29 \\
                                $M_\mathrm{disk}$ [$\%\,\mathrm{M}_\odot$]         & 5.58  & 9.4\\
                                $M_\mathrm{disk}$ [$\%\,M_\ast$]                        & 2.67  & 4.5 \\
                                $r_\mathrm{p}$ [au]                                             & 48      & 48 / 86 / 145\\
                                $\Sigma_{t=0}(r)$                       & $\propto r^{-3/2}$    & \filler \\
                                $\Sigma_\mathrm{p}$ [g/cm${}^2$]                        & 9.61  & 9.61 / 4.01 / 1.83\\
                                $h_\mathrm{p}$ [\%]                                             & 5.601 & 5.601 / 6.617 / 7.684\\
                                $T_\mathrm{p}$ [K]                                                      & 34.5  & 34.5 / 26.9 / 21.5\\
                                $\gamma$                                                                        & 1.4 & \filler \\
                                $\kappa$ [cm${}^2$/g]                                           & 0.45 & \filler \\
                                $\log\alpha$                                                            & $-4$ & \filler \\
                                $r_\mathrm{min}$--$r_\mathrm{max}$ [au] & 9.6--240 & 9.6--576 \\
                                $N_r\times N_\phi$                              & $573\times 1118$ & $800\times 1561$\\
                                $N_\mathrm{cells}/H_\mathrm{p}$                         & 10  & 11 / 13 / 15 \\
                        \end{tabular}           
                        \label{table:HD-multiplanet}
                \end{table}

        \end{appendix}
        
\end{document}